\documentclass[%
reprint,
superscriptaddress,
amsmath,amssymb,
pra,
]{revtex4-1}

\usepackage{graphicx}
\usepackage{hyperref}
\usepackage{dcolumn}
\usepackage{bm}
\usepackage{color}
\usepackage{soul}
 
\newcommand{\ket}[1]{\left| #1 \right>} 
\newcommand{\outprod}[2]{\left| #1 \vphantom{#2} \right>\left< #2 \vphantom{#1} \right|} 

\begin{document}
\preprint{APS/123-QED}

\title{Performance of dynamical decoupling in bosonic environments and under pulse-timing fluctuations}

\author{W. S. Teixeira}
\affiliation{Centro de Ci\^encias Naturais e Humanas, Universidade Federal do ABC, Santo Andr\'e, 09210-170 S\~ao Paulo, Brazil}
\author{K. T. Kapale}
\affiliation{Department of Physics, Western Illinois University, Macomb, Illinois 61455, USA}
\author{M. Paternostro}
\affiliation{Centre for Theoretical Atomic, Molecular and Optical Physics, School of Mathematics and Physics, Queen's University, Belfast BT7 1NN, United Kingdom}
\author{F. L. Semi\~ao}
\affiliation{Centro de Ci\^encias Naturais e Humanas, Universidade Federal do ABC, Santo Andr\'e, 09210-170 S\~ao Paulo, Brazil} 

%


\date{\today}

\begin{abstract}
We study the suppression of qubit dephasing through Uhrig dynamical decoupling (UDD) in nontrivial environments modeled within the spin-boson formalism. In particular, we address the case of (i) a qubit coupled to a bosonic bath with power-law spectral density, and (ii) a qubit coupled to a single harmonic oscillator that dissipates energy into a bosonic bath, which embodies an example of a structured bath for the qubit. We then model the influence of random time jitter in the UDD protocol by sorting pulse-application times from Gaussian distributions centered at appropriate values dictated by the optimal protocol. In case (i) we find that, when few pulses are applied and a sharp cutoff is considered, longer coherence times and robust UDD performances (against random timing errors) are achieved for a super-Ohmic bath. On the other hand, when an exponential cutoff is considered a super-Ohmic bath is undesirable. In case (ii) the best scenario is obtained for an overdamped harmonic motion.  
Our study provides relevant information for the implementation of optimized schemes for the protection of quantum states from decoherence. 
\end{abstract}
\maketitle
\section{\label{sec:level1}Introduction}
Quantum systems are highly susceptible to perturbations coming from imprecise experimental control and undesirable interactions with the surrounding environment. In general, system-environment coupling causes the vanishing of the {off}-diagonal matrix elements of the density operator for the system, wiping out its superposition aspects as time evolution occurs in a process called decoherence. The existence of such a phenomenon in realistic situations has motivated scientists and engineers to devise methods to protect quantum information, with quantum error correction and the use of decoherence-free subspaces as important examples~\cite{qec}. The technique of dynamical decoupling (DD) has emerged as a complementary and powerful means to combat decoherence experimentally. In general, it requires less physical resources than the {aforementioned} techniques and can, in principle, be used in a great variety of system-environment couplings.

In DD, a sequence of strong and short electromagnetic pulses is applied to the system in order to time reverse the effects of the system-bath interaction Hamiltonian. The idea stems from the spin-echo technique in liquid nuclear magnetic resonance (NMR) developed in 1950 by Hahn \cite{hahn50}. Sequences involving multiple pulses have been explored in NMR afterwards \cite{cp54,mg58,hae68} and incorporated into quantum information processing (QIP) schemes~\cite{vio98}. In 2007, Uhrig derived a sequence of unequally time-spaced pulses~\cite{uhr07} [dubbed Uhrig dynamical decoupling (UDD)], which efficiently suppresses decoherence and is more robust against temperature changes than any conventional DD schemes based on equidistant pulses. Even though UDD was originally proposed for qubit protection in the spin-boson model (SBM) under pure dephasing, also known as the independent boson model \cite{gil05,mah90}, it has been shown that UDD can be generalized to apply to other types of system-environment couplings~\cite{yang08}, time-dependent Hamiltonians~\cite{uhr08}, and concatenated to quench both spin relaxation and dephasing~\cite{wes10,uhr09}. UDD-based schemes have been successfully implemented in trapped ions \cite{bie09} and solid-state systems \cite{du09}. 

In this paper, we consider Uhrig's original scheme~\cite{uhr07} and apply it to nontrivial bath configurations with the goal of achieving the effective suppression of qubit decoherence. In particular, we study (i) an environment with a spectral density following the power law $\mathcal{J}(\omega)\propto \omega^s$ and with a cutoff frequency $\omega_c$; here, the exponent $s$ characterizes the dissipative dynamics; and (ii) (artificially) structured environments. In the so-called Ohmic version of case (i), i.e., for $s=1$, the environment-induced damping is linear, describing well systems such as Josephson flux qubits, metallic environments, and unidimensional phonons~\cite{wei08,lou95}. In the sub-Ohmic ($0<s<1$) and super-Ohmic ($s>1$) cases, the environmental effects become frequency dependent. Values of  $s=0.5$ are found in electron tunneling coupled to $RC$ transmission lines~\cite{wei08} and environments of nanomechanical devices~\cite{seo07}. Values of $s=3$ and $s=5$ can be found in defect tunneling in solids coupled to a three-dimensional bath of acoustic phonons \cite{wei08,sas90}.

Case (ii), which involves a structured environment, has been less explored in the context of decoherence suppression. The SBM here is used to describe a qubit coupled to a harmonic oscillator with frequency $\Omega$, which in turn is damped with a damping rate $\eta$ by a bath of harmonic oscillators \cite{wil04}. The effective spectral density $\mathcal{J}_{\text{eff}}(\omega)$ in this case has an Ohmic shape at low $\omega$, a peak that depends on the ratio between $\eta$ and $\Omega$, and a tailing behavior in the region of large values of $\omega$. Such a model was originally proposed in the context of electronic transport in biomolecules \cite{gar85} and has raised interest in condensed matter quantum computation as it describes the physics of a superconducting qubit coupled to stripline resonators~\cite{mar03}.

As the protection of quantum coherence via DD has become a prominent topic in practical realizations of QIP, our work also has the ambition to investigate how UDD is affected by plausible and important experimental problems such as time jitter of the control pulses. The origin of such perturbations is diverse and includes technical errors, noisy pumping sources, mechanical vibrations of the laser, and amplified spontaneous emission in mode-locked lasers~\cite{hje92,lu00}. 

The remainder of the paper is organized as follows. Section~\ref{model} describes the ideal UDD (iUDD) protocol, as well as our model to simulate randomness in the pulse-application times (pUDD). Section~\ref{results} discusses our results {achieved using pUDD for the case of a qubit directly coupled to a bosonic bath with power-law spectral density, whereas Sec.~\ref{results2} addresses the case of a qubit-structured environment interaction.} Finally, Sec.~\ref{conc} presents our conclusions.
\section{\label{model}The model}
The Hamiltonian of a qubit linearly coupled to a bath of harmonic oscillators that induces dephasing is given by \cite{wei08}
\begin{equation}
H = \frac{1}{2}\epsilon\sigma_z+\sum_{i}{\omega_i b_{i}^{\dagger}b_i}+\frac{1}{2}\sigma_z\sum_{i}{\lambda_i\left(b_{i}^{\dagger}+b_i\right)},
\label{eq:sbm}
\end{equation}
with $\sigma_z$ the $z$ Pauli matrix and $b_i$ ($b_i^\dagger$) the bosonic annihilation (creation) operators. We have chosen units such that $\hbar=1$ for ease of notation, and called $\epsilon$ the energy gap between the two logical states of the qubit, $\omega_i$ the frequency of the $i^{\text th}$ oscillator in the bath, and $\lambda_i$ its coupling strength to the qubit. In Eq.~\eqref{eq:sbm}, the bath is only coupled to $\sigma_z$, i.e., it just induces random phase changes in the qubit (dephasing), which is a common scenario in many experiments~\cite{leg87}. We have also neglected the possible tunneling mechanism between the states of the qubit by assuming the energy gap $\epsilon$ to be much larger than the tunneling rate. This is the case of electronic excitations in biomolecules interacting with a solvent environment~\cite{gil05}.

All relevant bath properties are contained in the spectral density \cite{wei08}
\begin{equation}
\mathcal{J}(\omega) = \sum_i \lambda_i^2\delta(\omega - \omega_i),
\label{eq:Jdef}
\end{equation}
whose explicit form depends on the type of system and environment being considered. In circuit quantum electrodynamics, information about $\mathcal{J}(\omega)$ can be obtained by inspecting either the effective damping or noise caused by an electronic circuit coupled to a superconducting qubit \cite{wil04}. In light-absorbing biomolecules, such as chromophores, $\mathcal{J}(\omega)$ can be extracted from ultra-fast laser spectroscopy \cite{gil08}. The form of Eq.~\eqref{eq:Jdef} involves a sequence of $\delta$ peaks at the frequencies of the oscillators in the bath. If the spectrum is dense, which is the case for a bath comprising a very large number of oscillators, the spectral density can be modeled as a continuous and smooth function up to some cutoff frequency $\omega_c$. A conventional assumption is that $\mathcal{J}(\omega)$ has a power-law behavior for small frequencies and vanishes in the limit $\omega\rightarrow\infty$ in order to avoid pathological phenomena \cite{leg87}. Thus, for this study, we use the spectral densities
\begin{eqnarray}
\mathcal{J}_{1}(\omega)=2\alpha \omega_c\left(\frac{\omega}{\omega_c}\right)^s \Theta(\omega_c-\omega), \label{eq:J1} \\
\mathcal{J}_{2}(\omega)=2\alpha \omega_c\left(\frac{\omega}{\omega_c}\right)^s e^{-\omega/\omega_c}, \label{eq:J2}
\end{eqnarray}
where $\alpha$ is the dimensionless effective coupling strength. The differences between $\mathcal{J}_{1}(\omega)$ and $\mathcal{J}_{2}(\omega)$ are solely due to the choices of the cutoff function. While the Heaviside distribution $\Theta$ determines a sharp cutoff for $\mathcal{J}_1(\omega)$, the exponential function establishes a smooth cutoff for $\mathcal{J}_2(\omega)$. As we shall see, the parameter $s$ and the cutoff function determine the robustness of UDD against random errors in the pulse-application times. This is one of our main results. 

The initial state of the qubit is set to be $\rho_q(0)=D_x(\pi/2)\outprod{\uparrow}{\uparrow}D^{\dagger}_x(\pi/2)$, with $\sigma_z\ket{\uparrow}=\ket{\uparrow}$ and $D_x(\phi)=\exp(-i\phi\sigma_x/2)$ being a rotation by an angle $\phi$ around the $x$ axis of the qubit Bloch sphere. Such an initial state is chosen due to its high coherence in the $\sigma_z$ basis. For the bosonic bath, we consider it to be in thermal equilibrium at temperature $T=1/\beta$ ($k_B=1$) so that its density matrix is $\rho_b(0)=\exp(-\beta H)/\text{Tr}\left[\exp(-\beta H)\right]$. We first consider the evolution with no dynamical decoupling, and the decoherence after an interaction time $t$ can be characterized by the mean value of the Pauli matrix $\sigma_y$ as~\cite{uhr07}
\begin{eqnarray}
r(t) &=& \text{Tr}\left[\sigma_y \rho(t)\right] = e^{-2\chi(t)},
\label{eq:s}
\end{eqnarray}
where
\begin{eqnarray}
\rho(t)=e^{-iHt}\left[\rho_q(0)\otimes \rho_b(0)\right]e^{iHt},
\label{eq:rho}
\end{eqnarray}
and
\begin{eqnarray}
\chi(t) = \int_{0}^{\infty}{\frac{\mathcal{J}(\omega)}{\omega^2}\sin^{2}{\left(\frac{\omega t}{2}\right)}\coth{\left(\frac{\beta \omega}{2}\right)}d\omega}.
\label{eq:chi}
\end{eqnarray}
In order to obtain the explicit expressions for $r(t)$ and $\chi(t)$, Eq.~\eqref{eq:sbm} must be diagonalized through the application of the spin-dependent displacement operator $\exp(\sigma_z K)$ with $K=\sum{[\lambda_i/(2\omega_i)](b_i^{\dagger}-b_i)}$. 

We now assess the evolution when the iUDD protocol proposed in Ref.~\cite{uhr07} is used. It consists of the application of $n$ instantaneous error-free $\pi$ pulses along the $y$ direction of the qubit Bloch sphere, modeled as $D_y(\pi)=i\sigma_y$ at instants of time $\delta_j \tau$. Here, $0<\delta_j<1$ and $\tau$ is the total evolution time. This procedure changes the signal in Eq.~\eqref{eq:s} to 
\begin{eqnarray}
r_n(\tau) &=& e^{-2\chi_n(\tau)},
\label{eq:r_n}
\end{eqnarray}
where
\begin{eqnarray}
\chi_n(\tau) = \int_{0}^{\infty}\frac{\mathcal{J}(\omega)}{4\omega^2}|y_n(\omega \tau)|^2\coth{\left(\frac{\beta \omega}{2}\right)}d\omega,
\label{eq:chi_n}
\end{eqnarray}
and
\begin{eqnarray}
y_n(\omega \tau)&=& 1+(-1)^{n+1}e^{i\omega \tau}+2\sum_{j=1}^{n}{(-1)^j e^{i \omega \delta_j \tau}}.
\label{eq:y_n}
\end{eqnarray}
The mathematical steps for obtaining Eqs.~\eqref{eq:r_n}-\eqref{eq:y_n} are detailed in Refs. \cite{uhr07,uhr08}. All pulse-sequence information is encoded in the function $|y_n(\omega \tau)|^2$ that, along with the integrand of $\chi_n(\tau)$, should be as close to zero as possible in order for the signal $r_n(\tau)$ to keep close to unity. For a given number of pulses $n$, such a condition occurs for
\begin{eqnarray}
\delta_j = \sin^2\left(\frac{\pi j}{2n+2}\right).
\label{eq:deltaj}
\end{eqnarray}
It is also worth keeping in mind that the approximation to instantaneous error-free $\pi$ pulses is theoretically convenient since the Hamiltonian of a single pulse is usually much stronger than the unperturbed system Hamiltonian. 

A natural question to ask is how random time jitter would influence the optimized decoherence suppression achieved by iUDD as the number of pulses increases. To model it, we evaluate the signal produced through Eq.~\eqref{eq:r_n} using perturbed pulse-application times
\begin{eqnarray}
\delta_{j,p}\tau = \delta_j\tau+a_j,
\label{eq:deltajp}
\end{eqnarray}
where $a_j$ is a parameter that varies randomly at each pulse according to a Gaussian distribution centered at $0$ and with a small standard deviation $\Lambda$. 
This condition is necessary to prevent $\delta_{j,p}$ from being negative or greater than unity as well as to obey $\delta_{j+1,p}>\delta_{j,p}$, which is another requirement to maintain the physical significance. 

By considering perturbations in $\delta_j \tau$ instead of  $\delta_j$, one guarantees  that all errors,  caused by some external source,  are uncorrelated and independent of the total elapsed time $\tau$. Furthermore, using a Gaussian distribution is a sensible choice given that the complete characterization of errors might not be straightforward in the laboratory or even theoretically. Other error models to the amplitude and phase of $\pi$ pulses have already been explored \cite{gea00,wan12,far15}.  However, the present paper explicitly considers a microscopic model for the system-bath arrangement and addresses the optimized sequences described by Eq.~\eqref{eq:deltaj} rather than periodic and concatenated sequences.

For our analysis, besides studying $1-\overline{r}_{n,p}(\tau)$, where $\overline{r}_{n,p}(\tau)$ is the average signal produced by pUDD after a large number of realizations, subjected to Gaussian fluctuations, we found it convenient to also define and study the quantity
\begin{eqnarray}
R_n(\tau)=1-\left|\overline{r}_{n,p}(\tau)-r_n(\tau)\right|.
\label{eq:rob}
\end{eqnarray}
Clearly, good experimental implementations of UDD would require values of $R_n(\tau)$ close to one because $\overline{r}_{n,p}(\tau)\rightarrow r_n$ as $\Lambda\rightarrow 0$, which corresponds to iUDD. In the next sections, we compare numerical results for qubit decoherence suppression produced by iUDD and pUDD in the configurations mentioned earlier.
\section{\label{results}UDD with power-law spectral density} 
\subsection{Sharp cutoff}
\begin{figure}[!h]
\centering
\includegraphics[scale=0.44]{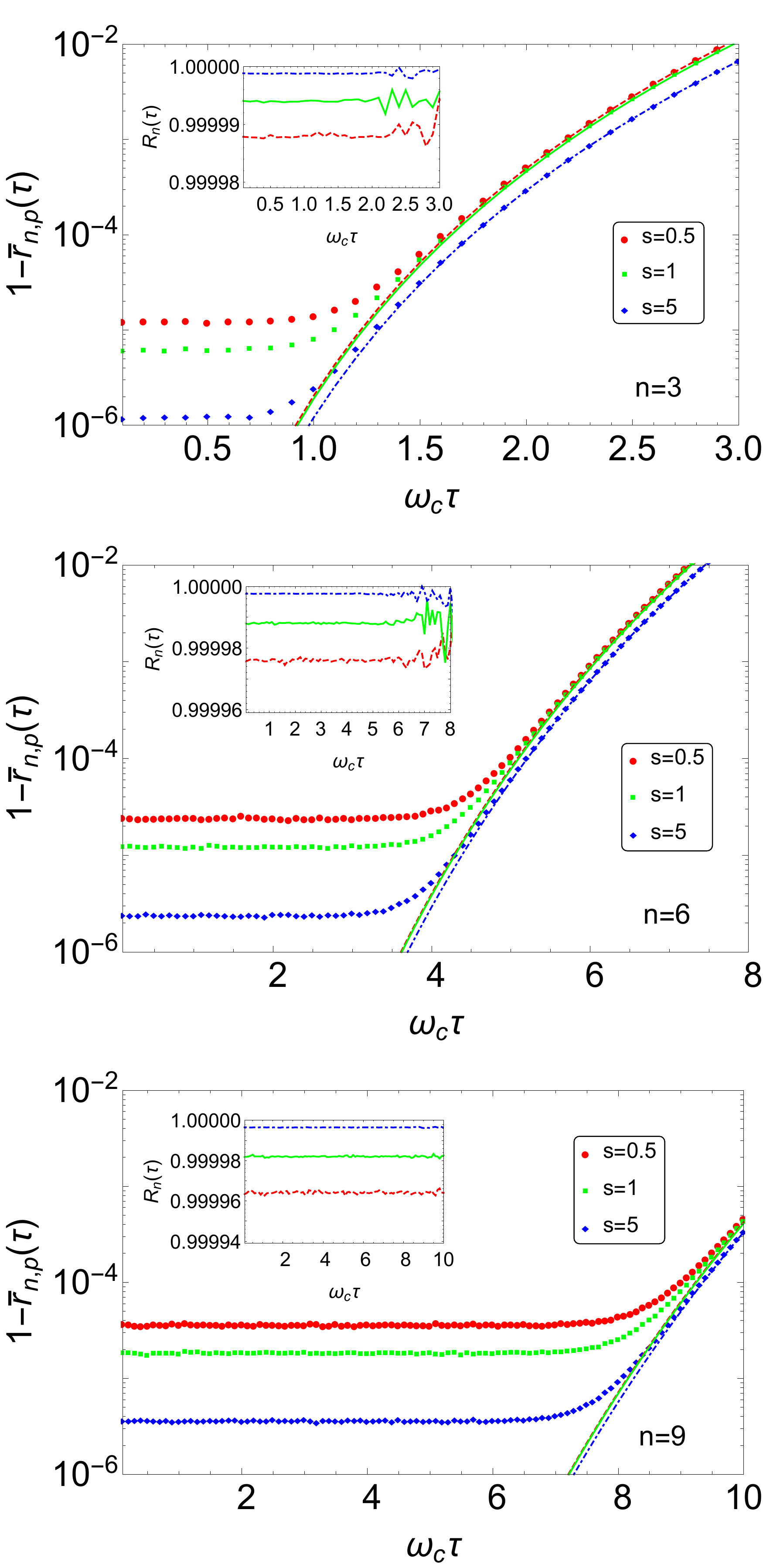}
\caption{\label{fg:comparison1} (Color online) Values of $1-\overline{r}_{n,p}(\tau)$ and $R_n(\tau)$ as functions of the dimensionless time $\omega_c \tau$ for different numbers of pulses $n$ and parameters $s$ in configuration (i) [with spectral density $\mathcal{J}_1(\omega)$]. Top panel: three pulses; middle panel: six pulses; and bottom panel: nine pulses. Dashed, solid, and dotted-dashed lines respectively represent iUDD for $s=0.5$, $s=1$, and $s=5$, whereas the corresponding markers represent pUDD. Each point in pUDD has been obtained through the mean value of $1-r_n(\tau)$ over $5000$ realizations and increment $0.1$ on $\omega_c \tau$. The chosen values of coupling strength, temperature, and Gaussian standard deviation are respectively $\alpha=0.1$, $T=10\ \omega_c$, and $\Lambda=5\times10^{-4}$.}
\label{fg:pUDD1}
\end{figure}

First, we carry out our analysis with a spectral density of the form described in Eq.~\eqref{eq:J1}. Figure~\ref{fg:pUDD1} shows the quantities $1-\overline{r}_{n,p}(\tau)$ and $R_n(\tau)$ for different numbers of pulses $n$ and choices of $s$. The fact that the deviations of pUDD from iUDD are numerically small is not fundamental  since this is just a consequence of the smallness of $\Lambda$ taken in the simulations. Our focus here is to characterize the deviations of pUDD from iUDD when $s$ and $n$ are varied.  As shown in Fig.~\ref{fg:pUDD1}, sub-Ohmic environments tend to be more affected by decoherence and are less robust against random timing errors. On the other hand, the super-Ohmic case is clearly more robust. A brief analysis of the integrand of $\chi_n(\tau)$ allows us to understand why smaller values of $s$ make UDD less efficient. First, for $\omega \tau /2 < n+1$, one can use the approximation \cite{uhr07}
\begin{eqnarray}
\left|y_n(\omega t)\right|^2 \approx 16(n+1)^2J_{n+1}^2\left(\frac{\omega \tau}{2}\right)
\label{eq:appBesselJ}
\end{eqnarray}
with $J_{n+1}$ the Bessel function of the first kind of order $n$. Second, by expanding $\coth\left({\beta\omega}/{2}\right)$ in the power series of $\beta\omega$ and using Eq.~\eqref{eq:J1}, $\chi_n(\tau)$ can be rewritten in terms of the dimensionless variable $\omega'=\omega/\omega_c$ as
\begin{equation}
\begin{aligned}
\chi_n(\tau)&=8(n+1)^2 \alpha \omega_c^{2}\biggl[\int_{0}^{1}\frac{2}{\beta \omega_c^{3}}\omega'^{s-3}J_{n+1}^2\left(\frac{\omega' \omega_c \tau}{2}\right)d\omega'  \\
&+\int_{0}^{1}\frac{\beta}{6\omega_c}\omega'^{s-1}J_{n+1}^2\left(\frac{\omega' \omega_c \tau}{2}\right)d\omega'  \\ 
&-\int_{0}^{1}\frac{\beta^3 \omega_c}{360}\omega'^{s+1}J_{n+1}^2\left(\frac{\omega' \omega_c \tau}{2}\right)d\omega' + \ldots \biggr].
\end{aligned}
\label{eq:appChi_n}
\end{equation}
It is clear now that smaller values of $s$ produce larger results for the integrals in Eq.~\eqref{eq:appChi_n} and, therefore, larger values of $\chi_n(\tau)$. This is precisely what is shown in  Fig.~\ref{fg:pUDD1}.  However, also according to Fig.~\ref{fg:pUDD1}, we see that the independence on $s$ for large number of pulses, a feature of iUDD, is not evident in pUDD for short times, when the random perturbations $a_j$ have more influence on the protocol. As a final remark concerning Eq.~\eqref{eq:appChi_n}, one can notice that $s\geq3$ prevents integrals from being divergent at $\omega'=0$, which makes super-Ohmic environments more robust against possible errors introduced by numerical integration. 

One may think of  $1-\overline{r}_{n,p}(\tau)$ as an indicator of tolerance or threshold for errors in units of memory (qubit). This is similar to what is normally done for quantum gates \cite{rei04,nil00}. Provided that errors are mainly introduced by random jitter, both threshold and protection time $\tau$ will play a fundamental role on the interpretation of the results presented here. According to Fig.~\ref{fg:pUDD1}, for a threshold set at $10^{-4}$ and the dimensionless protection time $\omega_c \tau \approx 1$, one has that $n=3$ pulses are already enough to keep the coherence of the qubit. However, for smaller thresholds, let us say $10^{-5}$ and again $\omega_c \tau \approx 1$, the application of three pulses is not useful for $s=0.5$, and the application of a large number of pulses, e.g., $n=6$, is even worse in the sense that just $s=5$ stays below the threshold. 
In other words, unlike iUDD, where a large number of pulses is always advantageous, this may not be true for pUDD. The reason is that each pulse contributes to errors in $\overline{r}_{n,p}(\tau)$ so that the application of more pulses causes larger deviations from the ideal signal. It is important to remark that Fig.~\ref{fg:pUDD1} reveals that devising a qubit interaction with a super-Ohmic bath is desirable for a power-law spectral density with a sharp cutoff, so that it can reduce the impacts of random timing jitter and produce better decoherence suppression as long as relatively few pulses are applied.
\subsection{Exponential cutoff}
\begin{figure}[!h]
\centering
\includegraphics[scale=0.44]{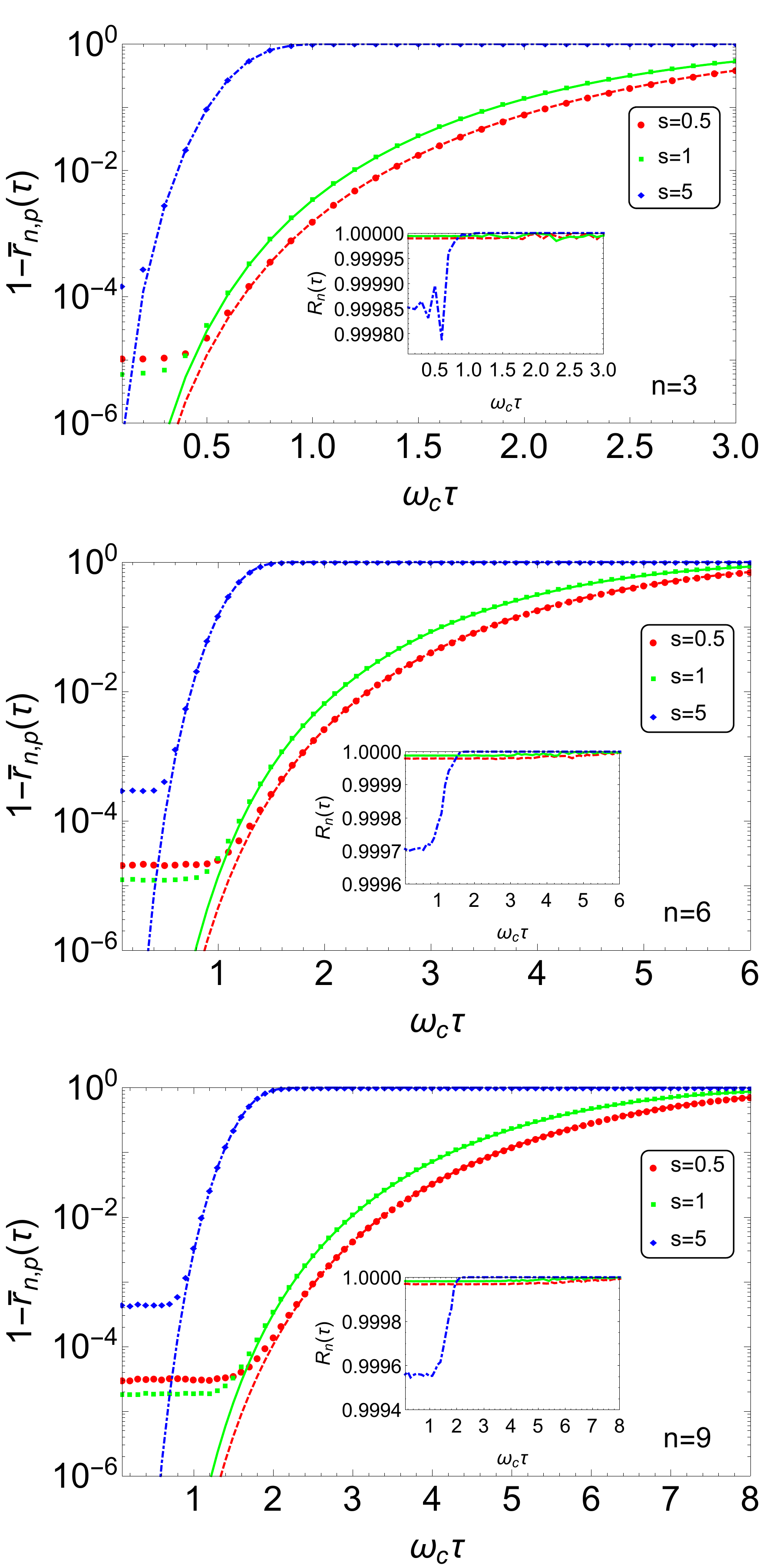}
\caption{\label{fg:comparison1b} (Color online) Values of $1-\overline{r}_{n,p}(\tau)$ and $R_n(\tau)$ as functions of the dimensionless time $\omega_c \tau$ for different numbers of pulses $n$ and parameters $s$ in configuration (i) [with spectral density $\mathcal{J}_2(\omega)$]. Top panel: three pulses; middle panel: six pulses; and bottom panel: nine pulses. Dashed, solid, and dotted-dashed lines respectively represent iUDD for $s=0.5$, $s=1$, and $s=5$, whereas the corresponding markers represent pUDD. All the other parameters are chosen as in Fig.~\ref{fg:pUDD1}.}
\label{fg:pUDD1b}
\end{figure}

\begin{figure}[!h]
\centering
\includegraphics[scale=0.35]{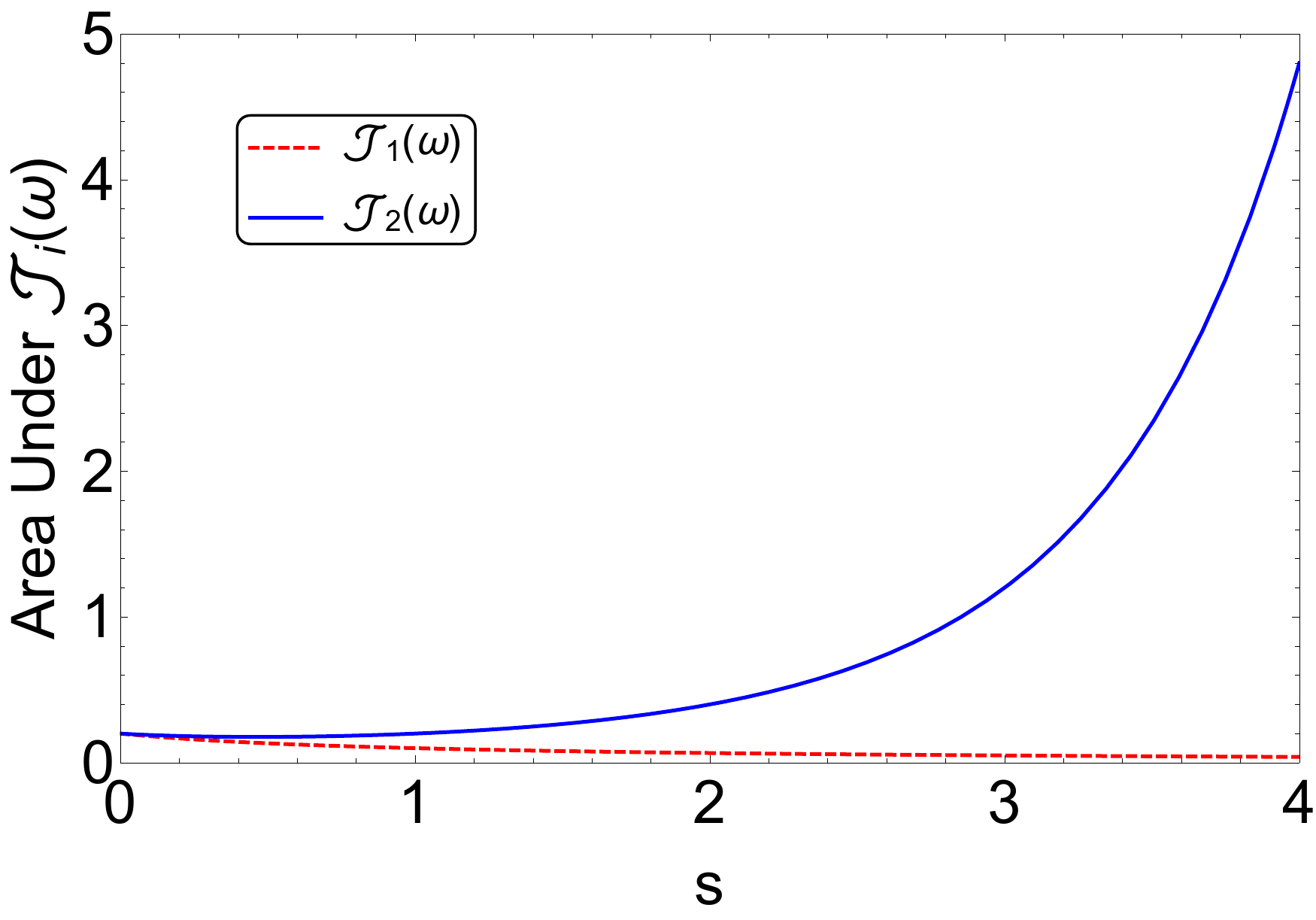}
\caption{\label{fg:comparison1b} (Color online) Areas under $\mathcal{J}_1(\omega)$ (solid line) and $\mathcal{J}_2(\omega)$ (dashed line) in arbitrary units for different values of $s$.}
\label{fg:ArJpl}
\end{figure}

We turn our analysis to the case where the power-law spectral density presents a smooth cutoff given by a decreasing exponential function as in Eq.~\eqref{eq:J2}. The comparison between $1-\overline{r}_{n,p}(\tau)$ and $R_n(\tau)$ for different numbers of pulses and $s$ is shown in Fig.~\ref{fg:pUDD1b}. The main point to notice is that the behavior of $1-\overline{r}_{n,p}(\tau)$ with respect to $s$ is opposite from the one observed in Fig.~\ref{fg:pUDD1}, so that the presence of a super-Ohmic bath makes the qubit more affected by decoherence and UDD sequences less robust against timing errors. The fact that the peaks of $\mathcal{J}_1(\omega)$ and $\mathcal{J}_2(\omega)$ are respectively at $\omega=\omega_c$ and $\omega=s\omega_c$ plays an important role here. For $\mathcal{J}_1(\omega)$, the integration in Eq.~\eqref{eq:chi_n} stops at $\omega=\omega_c$ and therefore the values of $s<1$ have larger contributions to $\chi_n(\tau)$. On the other hand, the integration in Eq.~\eqref{eq:chi_n} does not have an upper bound for $\mathcal{J}_2(\omega)$, which means that the values of $s>1$ contribute more to the deviation of the signal since the peak of the spectrum is beyond $\omega_c$. Indeed, the area under $\mathcal{J}_2(\omega)$ becomes considerably larger than the area under $\mathcal{J}_1(\omega)$ for super-Ohmic environments, as shown in Fig.~\ref{fg:ArJpl}. Further simulations involving Carr-Purcell-Meiboom-Gill (CPMG) sequences, defined by $\delta_j^{\text{CPMG}}=(j-1/2)/n$,  do not cause major changes to the robustness when compared to the ones presented in Fig.~\ref{fg:pUDD1b}, yet UDD has produced longer coherence times for small error thresholds. However, this might not be true for even smoother cutoff functions in the spectral density, e.g., $f(\omega')=1/(1+\omega'^2)$, where CPMG slightly outperforms UDD \cite{uhr08}. Nonetheless, in light of the explanation previously given, the increase in the number of pulses may lead to unwanted behavior in the sense of the maintenance of $1-\overline{r}_{n,p}(\tau)$ below a certain error threshold.
\section{\label{results2}UDD with structured environment}
In this section, we {discuss} case {(ii)}, where a qubit is coupled exclusively to a single damped harmonic oscillator $R$ with frequency $\Omega$. {The damping experienced by the oscillator arises due to its interaction with a bath.} This system is a paradigmatic model to represent the transfer of electrons between two localized sites in a biomolecule \cite{gar85}, as well as a qubit interaction with a prominent harmonic mode of a cavity field or vibration of a nano/microcantilever \cite{esc11,mil}. 

This complicated problem can be effectively converted to the simple spin-boson problem treated in configuration (i), where now the qubit is only subjected to an effective bosonic bath. What makes the problem interesting is that the interaction with the bath will be, in general, no longer governed by the power-law spectral densities of Eqs.~\eqref{eq:J1} and~\eqref{eq:J2}.  

The starting point is to consider an Ohmic spectral density for a bath in contact with the single harmonic oscillator $R$. This can be conveniently written as $\mathcal{J}_{\text{B}}(\omega)=2 \alpha \omega \exp(-\omega/\omega_c)$, with a cutoff frequency $\omega_c$. Then, with the use of normal coordinates for the bath, and the assumption of a broad continuum ($\omega_c\rightarrow \infty$), it is possible to make configuration (ii) assume the form of configuration (i), but with an effective spectral density $\mathcal{J}_{\text{eff}}(\omega)$ that reads \cite{gar85}
\begin{eqnarray}
\mathcal{J}_{\text{eff}}(\tilde{\omega})=\frac{2\alpha\tilde{\omega}\Omega}{(1-\tilde{\omega}^2)^2+4\tilde{\omega}^2\gamma^2},
\label{eq:JSE}
\end{eqnarray}
where $\tilde{\omega}=\omega/\Omega$ and  $\gamma=\eta/(2M\Omega)=\alpha/(M\Omega)$, with $M$ being the mass of the single harmonic oscillator $R$. Of course, if R is not a massive oscillator, there will be an equivalent definition for $\gamma$ in terms, for example, of electromagnetic constants and the volume of a cavity in the case of cavity quantum electrodynamics \cite{wal06}. The effective spectral density in Eq.~\eqref{eq:JSE} behaves Ohmically at low frequencies and goes to zero in the opposite limit. Also, contrary to the power-law case, it has a peak and this is both size and width dependent on $\gamma$. In underdamped cases ($\gamma \ll 1$), $\mathcal{J}_{\text{eff}}(\tilde{\omega})$ is narrow in the sense of a delta function and presents a peak around $\tilde{\omega}=1$. The greater values of $\gamma$ comprising the critical damping ($\gamma=1$) and overdamped cases ($\gamma \gg 1$) make $\mathcal{J}_{\text{eff}}(\tilde{\omega})$ broader with a peak that does not coincide with $\tilde{\omega}=1$.

Since the derivation of iUDD does not depend on a specific form of spectral density, we now can use Eq.~\eqref{eq:r_n} with $\mathcal{J}_{\text{eff}}(\tilde{\omega})$ to investigate the effects of coherence preservation under iUDD and random timing jitter under pUDD for configuration (ii). The important point here is that the qubit is again under pure dephasing, now caused by its interaction with a structured environment. This is an explicit investigation of iUDD for a structured environment, yet other optimized models of decoherence control in non-power-law spectra have been proposed \cite{gor08}. Even more interesting, the spectral density in Eq.~\eqref{eq:JSE} moves the problem to a context where the physical system can be a chromophore in a biomolecule \cite{gar85}. 

\begin{figure}[!h]
\centering
\includegraphics[scale=0.44]{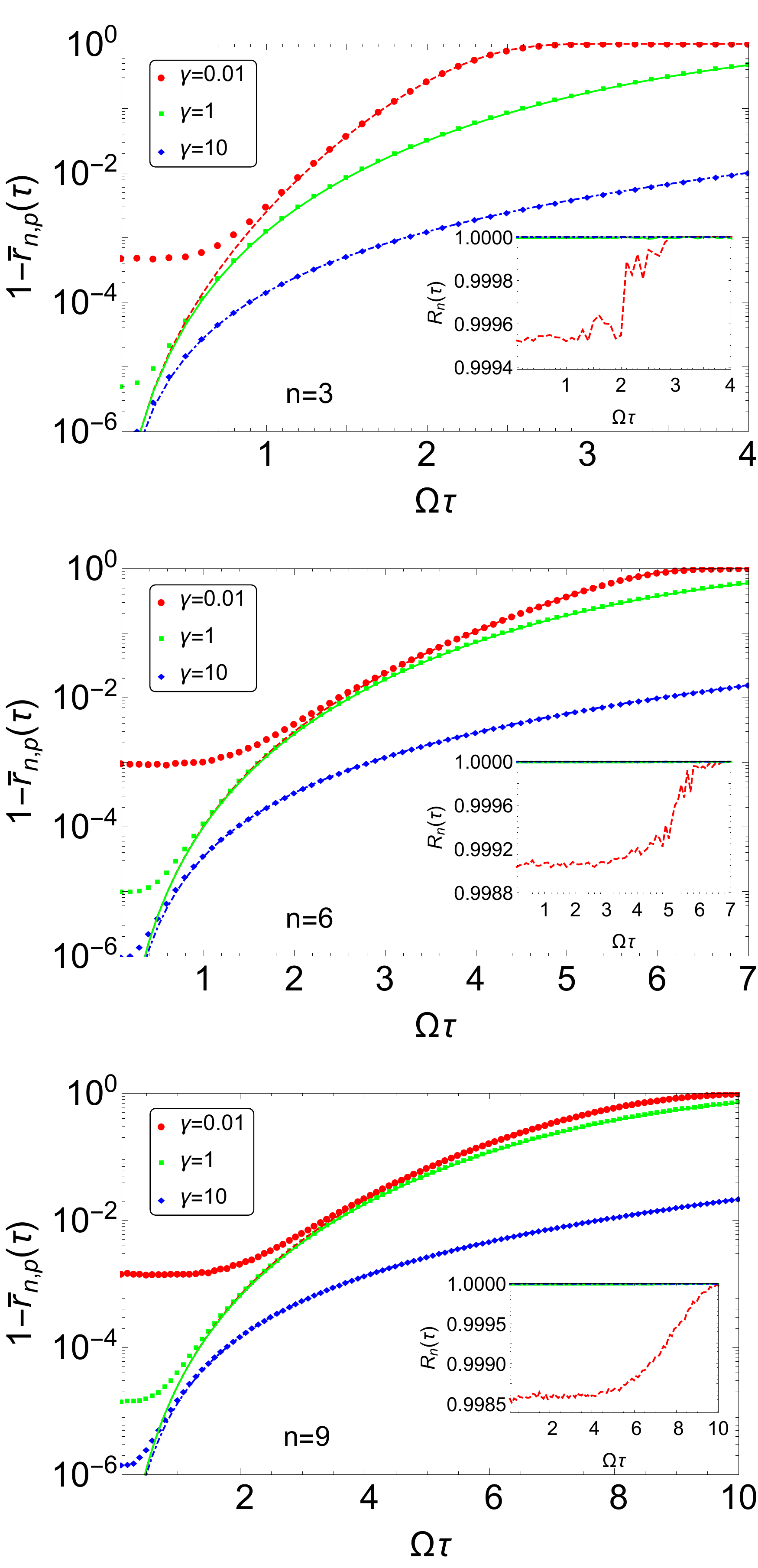}
\caption{\label{fg:comparison2} (Color online) Values of $1-\overline{r}_{n,p}(\tau)$ and $R_n(\tau)$ as functions of the dimensionless time $\Omega \tau$ for different number of pulses $n$ and parameters $\gamma$ in configuration (ii) [with spectral density $\mathcal{J}_{\text{eff}}(\omega)$]. Top panel: three pulses; middle panel: six pulses; and bottom panel: nine pulses. Dashed, solid, and dotted-dashed lines respectively represent iUDD for $\gamma=0.01$, $\gamma=1$, and $\gamma=10$, whereas the corresponding markers represent pUDD. Each point in pUDD has been obtained through the mean value of $1-r_n(\tau)$ over $5000$ realizations and increment $0.1$ on $\Omega \tau$. The chosen values of coupling strength, temperature, and Gaussian standard deviation are $\alpha=0.1$, $T=10\ \Omega$, and $\Lambda=5 \times 10^{-4}$.}
\label{fg:pUDD2}
\end{figure}

\begin{figure}[!h]
\centering
\includegraphics[scale=0.25]{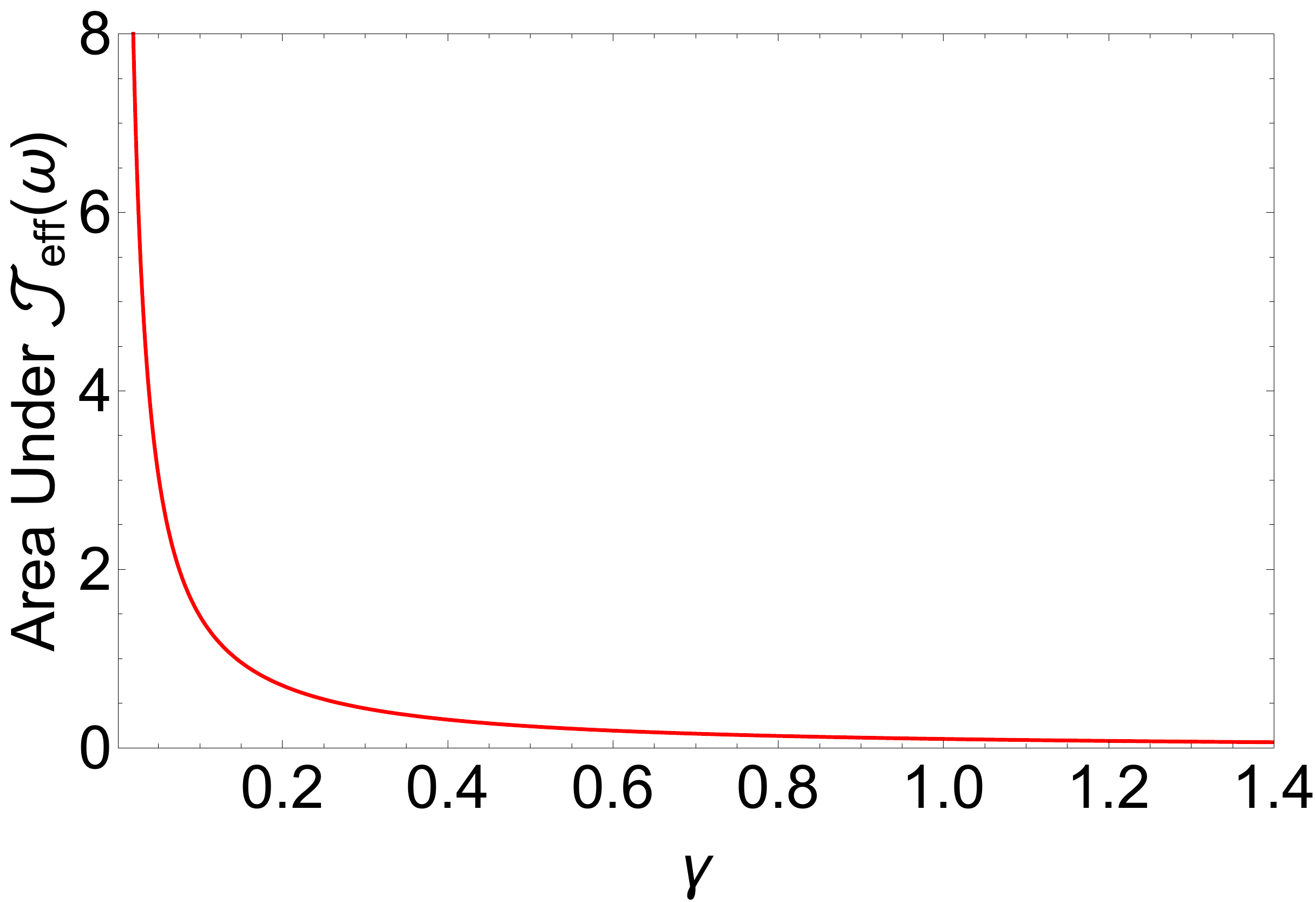}
\caption{\label{fg:comparison1b} (Color online) Area under $\mathcal{J}_{\text{eff}}(\omega)$ in arbitrary units for different values of $\gamma$.}
\label{fg:ArJse}
\end{figure}

For numerical integrations, the change of variable $\tilde{\omega}\rightarrow u/(1-u)$ has been applied to make the limits of integration finite in Eq.~\eqref{eq:chi_n}. In  Fig.~\ref{fg:pUDD2}, we compare decoherence suppression achieved by iUDD and pUDD using Eq.~\eqref{eq:JSE} for different numbers of pulses $n$ and values of $\gamma$. As before, solid lines represent iUDD, whereas each point representing pUDD is averaged over $5000$ realizations.
For short times ($\Omega\tau<1$), where fluctuations of $a_j$ are more decisive, $1-\overline{r}_{n,p}(\tau)$ reveals that underdamping is much more affected by timing jitter errors than critical and overdamping cases. The same can be seen from the behavior of $R_n(\tau)$ that stays substantially close to unity when oscillator $R$ is critical and overdamped, while this is not the case for underdamping. The explanation relies on the form of $\mathcal{J}_{\text{eff}}(\tilde{\omega})$ in Eq.~\eqref{eq:JSE}. Even though very small values of $\gamma$ promote narrow behaviors of $\mathcal{J}_{\text{eff}}(\tilde{\omega})$, they produce much greater peaks than large values of $\gamma$ do, and consequently larger areas under $\mathcal{J}_{\text{eff}}(\tilde{\omega})$ are produced (Fig.~\ref{fg:ArJse}). 
Therefore, small values of $\gamma$ make considerable contributions to $\chi_n(\tau)$ so that it becomes more sensitive to errors introduced by $\delta_{j,p}\tau$. We can then conclude that the presence of large damping in the oscillator $R$ is advantageous for coherence protection in the qubit and robustness against timing jitter errors when they are present.  Finally, from  Fig.~\ref{fg:pUDD2}, one can also see that larger numbers of pulses tend not to distinguish the values of $\gamma$ but, on the other hand, introduce more errors. The explanation is analogous to the one given for configuration (i).
\section{\label{conc}Conclusion}
We have studied the suppression of dephasing decoherence of a general qubit in contact with different types of dephasing environments that can be treated within the spin-boson model formalism. More specifically, we applied dynamical decoupling in the case of  (i) power-law spectral density and (ii) a structured environment consisting of a damped harmonic oscillator. We have focused on the optimized sequence of $\pi$ pulses presented in Ref. \cite{uhr07} and numerically included small noise to the pulse-application times in order to model the problem of timing jitter. Our results suggest that the error rate threshold, the chosen protection time, and the number of applied pulses dictate whether or not timing jitter can be neglected. Finally, we have seen that the application of a small number of pulses makes some regimes more attractive than others in the context of preservation of quantum coherence in iUDD and pUDD. In particular, for environments with power-law spectral density, the super-Ohmic regime is preferred just for a sharp cutoff function while for structured environments this is the case of overdamping. We believe the results presented here may find applications in contexts ranging from solid-state architectures to biomolecules in a solution. In particular, the application of dynamical decoupling to configuration (ii) may drive studies aimed at the preservation of quantum coherence and the demonstration of legitimate quantum effects, for instance, in chromophoric systems present in the photosynthetic apparatus of simple organisms. Dynamical decoupling applied to such systems may keep quantum coherence for longer times and favor the experimental assessment of nontrivial quantum effects.
\begin{acknowledgments}
W.S.T. would like to thank CAPES for a current scholarship and previous support by the ``Ci\^encia sem Fronteiras'' program. W.S.T. also acknowledges the hospitality of the Department of Physics, Western Illinois University, during the initial stages of this project.  K.T.K. would like to thank Tim Woodworth for fruitful discussions on the general topics of classical error correction and quantum information protection. M.P.  acknowledges  financial  support  from  John  Templeton  Foundation  (Grant  No. 43467), the EU Collaborative Project TherMiQ (Grant Agreement No. 618074), and  also gratefully acknowledges support from  the COST Action MP1209 
``Thermodynamics in the quantum regime." F.L.S. acknowledges support as a member of the Brazilian National Institute of Science and Technology of Quantum Information (INCT-IQ) and acknowledges partial support from CNPq 
(Grant No. 307774/2014-7).
\end{acknowledgments}

\end{document}